# Fast and Exact Top-k Search for Random Walk with Restart

Yasuhiro Fujiwara*‡, Makoto Nakatsuji†, Makoto Onizuka*, Masaru Kitsuregawa‡
*NTT Cyber Space Labs, †NTT Cyber Solutions Labs, ‡The University of Tokyo

{fujiwara.yasuhiro, nakatsuji.makoto, onizuka.makoto} @lab.ntt.co.jp, kitsure@tkl.iis.u-tokyo.ac.jp

## ABSTRACT

Graphs are fundamental data structures and have been employed for centuries to model real-world systems and phenomena. Random walk with restart (RWR) provides a good proximity score between two nodes in a graph, and it has been successfully used in many applications such as automatic image captioning, recommender systems, and link prediction. The goal of this work is to find nodes that have top-k highest proximities for a given node. Previous approaches to this problem find nodes efficiently at the expense of exactness. The main motivation of this paper is to answer, in the affirmative, the question, 'Is it possible to improve the search time without sacrificing the exactness?'. Our solution, *K-dash*, is based on two ideas: (1) It computes the proximity of a selected node efficiently by sparse matrices, and (2) It skips unnecessary proximity computations when searching for the top-k nodes. Theoretical analyses show that K-dash guarantees result exactness. We perform comprehensive experiments to verify the efficiency of K-dash. The results show that K-dash can find top-k nodes significantly faster than the previous approaches while it guarantees exactness.

## 1. INTRODUCTION

Recent advances in social and information science have shown that linked data pervade our society and the natural world around us [24]. Graphs become increasingly important to represent complicated structures and schema-less data such as is generated by Wikipedia [1], Freebase [2], and various social networks [10]. Due to the extensive applications of graph models, vast amounts of graph data have been collected and graph databases have attracted significant attention in the database community. In recent years, various approaches have been proposed to deal with graph-related research problems such as subgraph search [24], shortest-path query [4], pattern match [5], and graph clustering [25] to get insights into graph structures.

[1] http://www.wikipedia.org/
[2] http://www.freebase.com/

With the rapidly increasing amounts of graph data, searching graph databases to identify high proximity nodes, where a proximity measure is used to rank nodes in accordance with relevance to a query node [13], has become an important research problem. Many papers in the database community have addressed node-to-node proximities [20, 1, 13]. For example, Sun et al. proposed a novel proximity measure called PathSim which produces good similarity qualities given heterogeneous information networks [20]. Simrank++, proposed by Antonellis et al., finds high proximity nodes effectively for historical click data [1]. One of the most successful techniques known to the academic communities is based on *random walk with restart (RWR)* [19]. This is because the proximity defined by RWR yields the following benefits: (1) it captures the global structure of the graph [8], and (2) it captures multi-facet relationships between two nodes unlike traditional graph distances [21].

However, the computation of the proximities by RWR is computationally expensive. Consider a random particle that starts from query node $q$. The particle iteratively moves to its neighborhood with the probability that is proportional to their edge weights. Additionally, in each step there is a probability that it will restart at node $q$. A node probability changes over time during iterations by recursively applying the above procedures. As the result, the steady-state probability can be obtained. The proximity of node $u$ with respect to node $q$ is defined as the steady-state probability with which the particle stays at node $u$.

Although RWR has been receiving increasing interests from many applications [15, 11, 12, 19], its excessive CPU time led to the introduction of approximate approaches [22, 19]. These approaches have the advantage of speed at the expense of exactness. However, approximate algorithms are not well adopted. This is because it is difficult for approximate algorithms to enhance the quality of real applications. Therefore, we address the following problem in this paper:

**Problem** (Top-k search for RWR).

**Given:** *The query node $q$, and the required number of answer nodes $K$.*

**Find:** *Top $K$ nodes with the highest proximities with respect to node $q$ exactly.*

To the best of our knowledge, our approach to finding top-k nodes in RWR is the first solution to achieve both exactness and efficiency at the same time.

### 1.1 Contributions

We propose a novel method called *K-dash* that can efficiently find top-k nodes in RWR. In order to reduce search cost, (1) we use sparse matrices to compute the exact proximity of a selected node, and (2) we prune low proximity nodes by estimating the proximities of those nodes without computing their exact proximities. K-dash has the following attractive characteristics based on the above ideas:

- **Exact:** K-dash does not sacrifice accuracy even though it exploits an estimation-based approach to prune unlikely nodes; it returns top-k nodes without error unlike the previous approximate approaches.

- **Efficient:** K-dash practically requires $O(n+m)$ time where $n$ and $m$ are the number of nodes and edges, respectively. By comparison, solutions based on existing approximate algorithms are expensive; they need $O(n^2)$ time to find the answer nodes. Note that $m \ll n^2$ in practice [14].

- **Nimble:** K-dash practically needs $O(n+m)$ space while the previous approaches need $O(n^2)$ space. The required memory space of K-dash is smaller than that of the previous approximate approaches.

- **Parameter-free:** The previous approaches require careful setting of the inner-parameter [22], since it impacts the search results. K-dash, however, is completely automatic; this means it does not require the user to set any inner-parameters.

While RWR has been used in many applications, it has been difficult to utilize it due to its high computational cost. However, by providing exact solutions in a highly efficient manner, K-dash will allow many more RWR-based applications to be developed in the future.

The remainder of this paper is organized as follows. Section 2 describes related work. Section 3 overviews the background of this work. Section 4 introduces the main ideas of K-dash. Section 5 gives theoretical analyses of K-dash. Section 6 reviews the results of our experiments. Section 7 provides our brief conclusion.

## 2. RELATED WORK

Node-to-node proximity is an important property. One of the most popular proximity measurements is RWR, and researchers of data engineering have published many papers on RWR and its applications [15, 11, 12, 19, 22]. With our approach, many applications can be processed more efficiently.

*Application.* Automatic image captioning is a technique which automatically assigns caption words to a query image. Pan et al. proposed a graph-based automatic caption method in which images and caption words are treated as nodes in a mixed media graph [15]. They utilized RWR to estimate the correlations between the query image and the captions. They reported that their method provided 10 percent higher captioning accuracy than a fine-tuned method.

Recommendation systems aim to provide personalized recommendations of items to a user. One recent recommendation technique proposed by Konstas et al. is based on RWR over a graph that connects users to tags and tags to items, where the probabilities of relevance for items are given by RWR proximities; high interest items would have high proximities. They incorporate the additional information such as friendship and social tagging embedded in social knowledge to improve the accuracy of item recommendations [11]. They also applied a standard collaborative filtering method as a baseline, and showed that their method was superior.

The question of 'which new interactions among social network members are more likely to occur in the near future?' is being avidly pursued by many researchers. Schifanella et al. proposed a metadata based approach for this problem [17]. Their idea is that members with similar interests are more likely to be friends, so semantic similarity measures among members based on their annotation metadata should be predictive of social links. Liven-Nowell et al. explored this question by using RWR [12]; the probability of a future collaboration between authors is computed from RWR proximity. Their approach is based on the observation that the topology of the network can suggest many new collaborations. For example, two researchers who are close in the network will have many colleagues in common, and thus are more likely to collaborate in the near future. They took the RWR-based approach since it can capture the global structure of the graph. They showed that RWR provides better link predictions than the random prediction approach.

*Approximation method.* Even though RWR is very useful, one problem is its large CPU time. Sun et al. observed that the distribution of RWR proximities is highly skewed. Based on this observation, combined with the factor that many real graphs have block-wise/partition structure, they proposed an approximation approach for RWR [19]; they performed RWR only on the partition that contains the query node. All nodes outside the partition are simply assigned RWR proximities of 0. In other words, their approach outputs a local estimation of RWR proximities.

Tong et al. proposed a fast approximation solution for RWR. They designed B_LIN and its derivative, NB_LIN [22]. These methods take advantages of the block-wise structure and linear correlations in the adjacency matrix of real graphs, using the Sherman-Morrison Lemma [16] and the singular value decomposition (SVD). Especially for NB_LIN, they showed the proof of an error bound. The experimental results showed that their methods outperformed the approximation method of Sun et al. [19]. Their methods require $O(n^2)$ space and $O(n^2)$ time. This is because their methods utilize $O(n^2)$ size matrices to approximate the adjacency matrix for proximity computation.

## 3. PRELIMINARY

In this section, we formally define the notations and introduce the background of this paper. Table 1 lists the main symbols and their definitions.

Measuring the proximity of two nodes in a graph can be achieved using RWR. Starting from a node, a RWR is performed by iteratively following an edge to another node at each step. Additionally, at every step, there is a non-zero probability, $c$, of returning to the start node. Let **p** be a column vector where $p_u$ denotes the probability that the random walk is at node $u$. **q** is a column vector of zeros with the element corresponding to the starting node $q$ set to 1, i.e. $q_q = 1$. Also let **A** be the column normalized adjacency matrix of the graph. In other words, **A** is the transition probability table where its element $A_{uv}$ gives the probability of node $u$ being the next state given that the current state is node $v$. The steady-state, or stationary probabili-

443

Table 1: Definition of main symbols.

| Symbol | Definition |
|---|---|
| $q$ | query node |
| $K$ | Number of answer nodes |
| $n$ | Number of nodes |
| $m$ | Number of edges |
| $c$ | the restart probability |
| **p** | $n \times 1$ vector, $p_u$ is the proximity of node $u$ |
| **q** | $n \times 1$ vector, the $q$-th element 1 and 0 for others |
| **A** | the column normalized adjacent matrix |

ties for each node can be obtained by recursively applying the following equation until convergence:

$$\mathbf{p} = (1-c)\mathbf{A}\mathbf{p} + c\mathbf{q} \qquad (1)$$

where the convergence of the equation is guaranteed [18]. The steady-state probabilities give us the long term visit rate of each node given a bias toward query node $q$. Therefore, $p_u$ can be considered as a measure of proximity of node $u$ with respect to node $q$.

This method needs $O(mt)$ time where $t$ is the number of iteration steps. This incurs excessive CPU time for large graphs, and a fast solution is demanded as illustrated by the statement 'its on-line response time is not acceptable in real life situations' made in a previous study [11]. It should be emphasized that shortening response time is critical to enhancing business success in real web applications [3].

## 4. PROPOSED METHOD

In this section, we explain the two main ideas underlying K-dash. The main advantage of our approach is to exactly and efficiently find top-k highest proximity nodes for RWR. First, we give an overview of each idea and then a full description. Proofs of lemmas or theorems in this section are shown in Appendix A.

### 4.1 Ideas behind K-dash

Our solution is based on the following two approaches:

*Sparse matrices computation.* The proximities for a query node are the steady-state probabilities which are computed by recursive procedures as described in Section 3. This approach requires high computation time because it computes the proximities of all the nodes in the graph. Our idea is simple; we compute the proximities of only selected nodes enough to find the top-k nodes, instead of computing the proximities of all nodes.

As described in Section 4.2.1, the proximities of selected nodes are naively computed by the inverse matrix that can be directly obtained from Equation (1). Therefore, if we precompute and store this inverse matrix, we can get the proximities efficiently. However, this approach is impractical when the dataset is large, because it requires quadratic space to hold the inverse matrix.

We introduce an efficient approach that can compute the proximities from sparse matrices. In the precomputing process, we reorder nodes and compute the inverse matrices of the lower/upper triangulars obtained by LU decomposition as described in Section 4.2.2. A lower/upper triangular matrix is a matrix where all the elements above/below the main diagonal are zero. As a result, the inverse matrices are sparse, and we can compute the proximities of the selected nodes with low memory consumption by using the adjacency-list representation [6].

---

[3] http://www.keynote.com/downloads/Zona_Need_For_Speed.pdf

This new idea has the following two major advantages besides the one described above. First we can compute the proximities exactly. This is because LU decomposition, unlike SVD which is used in the previous methods [22], is not an approximation method. The second advantage is that we can compute the proximities efficiently. This is because we use sparse matrices to compute the proximities.

*Tree estimation.* Although our sparse matrices approach is able to compute the proximities of selected nodes, we have the following two questions to find the top-k nodes: (1) 'What nodes should be selected to compute the proximities in the search process?', and (2) 'Can we avoid computing the proximities of unselected nodes?'. The second approach is designed to answer these two questions.

These questions can be answered by estimating what nodes can be expected to have high/low proximities. Our proposal exploits the following observations: the proximity of a node declines as the number of hops from the query node increases, and proximities of unselected nodes can be estimated from computed proximities. Our search algorithm first constructs a single breadth-first search tree rooted at the query node. We compute the proximities of the top-k nearest nodes from the root node to discover answer candidate nodes. We then estimate the proximities of unselected nodes from the proximities of already selected nodes to obtain the upper proximity bound. The time incurred to estimate node proximity is $O(1)$ for each node. In the search process, if the upper proximity bound of a node gives a score lower than the $K$-th highest proximity of the candidates nodes, the node cannot be one of the top-k highest proximity nodes. Accordingly, unnecessary proximity computations can be skipped.

This estimation allows us to find the top-k nodes exactly while we prune unselected nodes. This means we can safely discard unlikely nodes at low CPU cost. This estimation approach also allows us to automatically determine the nodes for which we compute the proximities. This implies our approach avoids to have user-defined inner-parameters.

### 4.2 Sparse matrices computation

Our first approach is to obtain sparse inverse matrices to compute the proximities of selected nodes efficiently. In this section, we first describe how to compute the proximities by inverse matrices. We then describe that obtaining the sparse inverse matrices is an **NP**-complete problem, and we then show our approximate approach for the problem.

#### 4.2.1 Proximity computation

From Equation (1), we can obtain the following equation:

$$\mathbf{p} = c\{\mathbf{I} - (1-c)\mathbf{A}\}^{-1}\mathbf{q} = c\mathbf{W}^{-1}\mathbf{q} \qquad (2)$$

where **I** represents the identity matrix and $\mathbf{W} = \mathbf{I} - (1-c)\mathbf{A}$. This equation implies that we can compute the proximities of selected nodes by obtaining the corresponding elements in the inverse matrix $\mathbf{W}^{-1}$. However, this approach requires high memory consumption. This is because the inverse matrix $\mathbf{W}^{-1}$ would be dense even though the matrix $\mathbf{W}$ itself is sparse [16] (In many real graphs, the number of edges is much smaller than the squared number of nodes [14]). That is, this approach requires $O(n^2)$ space.

We utilize the inverse matrices of lower/upper triangulars to compute the proximities in our approach. Formally, the following equation gives the proximities for the query node, where the matrix $\mathbf{W}$ is decomposed to $\mathbf{LU}$ by the LU decomposition (i.e. $\mathbf{W} = \mathbf{LU}$).

$$\mathbf{p} = c\mathbf{U}^{-1}\mathbf{L}^{-1}\mathbf{q} \qquad (3)$$

Note that the matrices $\mathbf{L}^{-1}$ and $\mathbf{U}^{-1}$ are lower and upper triangular, respectively.

### 4.2.2 Inverse matrices problem

As shown in Equation (3), if we precompute the matrices $\mathbf{L}^{-1}$ and $\mathbf{U}^{-1}$, we can compute proximities of selected nodes. However, this raises the following question: 'Can the matrices $\mathbf{L}^{-1}$ and $\mathbf{U}^{-1}$ be sparse if matrix $\mathbf{W}^{-1}$ is dense?'. Our answer to this question is to compute the sparse matrices $\mathbf{L}^{-1}$ and $\mathbf{U}^{-1}$ by reordering the columns and rows of the sparse matrix $\mathbf{A}$. But finding the node order in matrix $\mathbf{A}$ that yields the sparse matrices is **NP**-complete.

**Theorem 1** (INVERSE MATRICES PROBLEM).
*Determining the node order that minimizes non-zero elements in matrices $\mathbf{L}^{-1}$ and $\mathbf{U}^{-1}$ is **NP**-complete.*

Because the INVERSE MATRICES problem is **NP**-complete, we introduce an approximation to address this problem. Before we describe our approaches in detail, we show the matrix elements of $\mathbf{L}^{-1}$ and $\mathbf{U}^{-1}$ can be represented by those of $\mathbf{L}$ and $\mathbf{U}$ by forward/backward substitution [16] as follows:

$$L^{-1}_{ij} = \begin{cases} 0 & (i < j) \\ 1/L_{ij} & (i = j) \\ -1/L_{ii} \sum_{k=j}^{i-1} L_{ik} L^{-1}_{kj} & (i > j) \end{cases} \qquad (4)$$

$$U^{-1}_{ij} = \begin{cases} 0 & (i > j) \\ 1/U_{ij} & (i = j) \\ -1/U_{ii} \sum_{k=i+1}^{j} U_{ik} U^{-1}_{kj} & (i < j) \end{cases} \qquad (5)$$

where the matrix elements of $\mathbf{L}$ and $\mathbf{U}$ can be represented by those of $\mathbf{W}$ by Crout's algorithm [16] as follows:

$$L_{ij} = \begin{cases} 0 & (i < j) \\ 1 & (i = j) \\ 1/U_{jj} \left( W_{ij} - \sum_{k=1}^{j-1} L_{ik} U_{kj} \right) & (i > j) \end{cases} \qquad (6)$$

$$U_{ij} = \begin{cases} 0 & (i > j) \\ W_{ij} & (i \leq j \cap i = 1) \\ W_{ij} - \sum_{k=1}^{i-1} L_{ik} U_{kj} & (i \leq j \cap i \neq 1) \end{cases} \qquad (7)$$

Equation (4), (5), (6), and (7) imply that elements of $\mathbf{L}^{-1}$, $\mathbf{U}^{-1}$, $\mathbf{L}$, and $\mathbf{U}$ are computed from the columns from left to right, and within each column from top to bottom. For example, element $L^{-1}_{ij}$ can be computed from the corresponding upper/left elements of $\mathbf{L}$ and $\mathbf{L}^{-1}$, and element $L_{ij}$ can be computed from the corresponding upper/left elements of $\mathbf{W}$, $\mathbf{L}$, and $\mathbf{U}$.

Our approaches are based on the following three observations in the above four equations: (1) elements $L^{-1}_{ij}$ and $U^{-1}_{ij}$ would be zero if the corresponding upper/left elements of $\mathbf{L}$ and $\mathbf{U}$ are zero, (2) upper/left elements of $\mathbf{L}$ and $\mathbf{U}$ would be zero if the corresponding upper/left elements of $\mathbf{W}$ are zero, and (3) the upper/left elements of $\mathbf{W}$ would be zero if the corresponding upper/left elements of $\mathbf{A}$ are zero since $\mathbf{A} = (\mathbf{I} - \mathbf{W})/(1 - c)$. That is, we can effectively compute

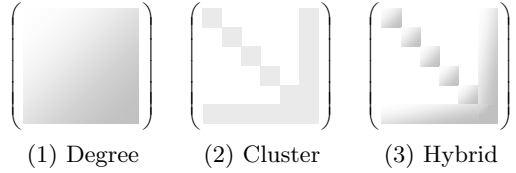

(1) Degree  (2) Cluster  (3) Hybrid

**Figure 1: Reordering methods.**

the sparse inverse matrices by letting upper/left elements of matrix $\mathbf{A}$ be zero.

We introduce the following three approximation solutions against the INVERSE MATRICES problem:

*Degree reordering.* In this approach, we arrange nodes of the given graph in ascending order of degree (the number of edges incident to a node) and rename them by the order. Low degree nodes have few edges, and the upper/left elements of corresponding matrix $\mathbf{A}$ are expected to be 0 with this approach.

*Cluster reordering.* This approach first divides the given graph into $\kappa$ partitions by Louvain Method [3], and it arranges nodes according to the partitions. Note that the number of partitions, $\kappa$, is automatically determined by Louvain Method. It then creates new empty $\kappa+1$-th partition. Finally, if a node of a partition has an edge to another partition, it rearranges the node to the $\kappa+1$-th partition. As a result, matrix $\mathbf{A}$ would be a doubly-bordered block diagonal matrix [16] as shown in Figure 1-(2); elements correspond to cross-partitions edges would be 0 for $\kappa$ partitions [4].

We use the Louvain Method because it is an efficient approach [5] and it utilizes the modularity [3] as the quality measure for partitioning. The modularity assesses the fitness of node partitioning, in the sense that there are many edges within a partition and only a few between them. That is, Louvain Method reduces the number of cross-partition edges. As a result, this approach should yield more sparse inverse matrices.

*Hybrid reordering.* This approach is a combination of the degree and the cluster reordering. That is, we arrange nodes by cluster reordering, and we then sort nodes in each partition by their degree. This approach makes matrix $\mathbf{A}$ have no cross-partition edges for $\kappa$ partitions, and the upper/left elements of each partition are expected to be 0.

Figure 1 illustrates matrix $\mathbf{A}$ obtained by above each approximation approach for the INVERSE MATRICES problem, where zero and non-zero elements are shown in white and gray, respectively. Algorithm 1, 2, and 3 in Appendix B show the details of degree reordering, cluster reordering, and hybrid reordering, respectively.

Owing to these three approaches, we can effectively obtain sparse matrices $\mathbf{L}$ and $\mathbf{U}$, and then sparse matrices $\mathbf{L}^{-1}$ and $\mathbf{U}^{-1}$. As demonstrated in Section 6, these approaches can drastically reduce the memory needed to hold matrices $\mathbf{L}^{-1}$ and $\mathbf{U}^{-1}$; they have practically linear space complexity in the size of edges in the given graph by using the adjacency-list representation [6].

---
[4]Doubly-bordered block diagonal matrix $D$ is defined by $D_{uv} = 0, \forall\{(u,v)|P(u) \neq P(v), P(u) \neq \kappa + 1, P(v) \neq \kappa + 1\}$ where $P(u)$ is the partition number of node $u$.
[5]For all data in our experiments, Louvain Method can compute partitions in a few seconds.

## 4.3 Tree estimation

We introduce an algorithm for estimating the proximities of unselected nodes in the search process effectively and efficiently. In this approach, a node is visited one by one, and we estimate its proximity. If the estimated proximity is not lower than the $K$-th highest proximity of candidate nodes, then the node is selected to compute the exact proximity. Otherwise we skip subsequent exact proximity computations of visited nodes. This approach, based on a single breadth-first search tree, yields the upper bounding score estimations of visited nodes. In this section, we first give the notations for the estimation, next we formally introduce the estimation, and then our approach to incremental estimation in the search process.

### 4.3.1 Notation

In the search process, we construct a single breadth-first search tree that is rooted on the query node; thus it forms layer 0. The direct neighbors of the root node form layer 1. All nodes that are $i$ hops from the root node form layer $i$.

The set of nodes in the graph is defined as $V$, and the set of selected (i.e., exact proximity computed) nodes is defined as $V_s$. The layer number of node $u$ is denoted as $l_u$. Moreover, the set of selected nodes prior node $u$ whose layer number is $l_u$ is defined as $V_{l_u}(u)$, that is $V_{l_u}(u) = \{v : (v \in V_s) \cap (l_v = l_u)\}$. $A_{max}$ is the maximum element in matrix $\mathbf{A}$, that is $A_{max} = \max\{A_{ij} : i, j \in V\}$. $A_{max}(u)$ is the maximum element from node $u$, that is $A_{max}(u) = \max\{A_{iu} : i \in V\}$.

Note that both $A_{max}$ and $A_{max}(u)$ can be precomputed. It requires $O(1)$ space to hold $A_{max}$, and it requires $O(n)$ space to hold $A_{max}(u)$ of all $n$ nodes.

### 4.3.2 Proximity estimation

We describe the definition of the proximity estimation in this section. We also show that the estimation gives a valid upper proximity bound. We estimate the proximity of node $u$ via breadth-first search tree as follows:

**Definition 1** (PROXIMITY ESTIMATION). *If node $u$ is not the query node (i.e. $u \neq q$), the following equation gives the definition of proximity estimation of node $u$, $\bar{p}_u$, to skip proximity computation in the search process:*

$$\bar{p}_u = c' \left\{ \sum_{v \in V_{l_u-1}(u)} p_v A_{max}(v) + \sum_{v \in V_{l_u}(u)} p_v A_{max}(v) + \left(1 - \sum_{v \in V_s} p_v\right) A_{max} \right\} \quad (8)$$

*where $c' = (1-c)/(1 - A_{uu} + cA_{uu})$.*
*If node $u$ is the query node (i.e. $u = q$), $\bar{p}_u = 1$.*

It needs $O(n)$ time to compute the estimation for each node if we compute it according to Definition 1. This is because $V_{l_u-1}(u)$, $V_{l_u}(u)$, and $V_s$ all have size of $O(n)$. We, however, compute the estimation in $O(1)$ time as described in Section 4.3.3.

To show the property of our proximity estimation, we introduce the following lemma:

**Lemma 1** (PROXIMITY ESTIMATION). $\bar{p}_u \geq p_u$ *holds for node $u$ in the given graph.*

This lemma enables us to find the answer nodes exactly.

The determination of the root node of the tree and the selection of proximity computation nodes are important in achieving efficient pruning. We determine the query node as the root node, and we visit and select nodes in increasing order of layer number. This is because: (1) a few nodes which are just a few hops from the query node have high proximities, and (2) we can estimate the proximities of visited nodes from those of selected nodes (see Definition 1). As a result, we can effectively estimate the proximities of visited nodes.

If the estimated proximity of a visited node is lower than the $K$-th highest proximity of the candidate nodes, we terminate the search process without computing the estimations of other unvisited and unselected nodes. However, this raise the following question: 'Can we find the answer nodes exactly if we terminate the search process?'. To answer this question, we show the following lemma:

**Lemma 2** (LAYER SEARCH). *If nodes are visited and selected in ascending order of layers, $\bar{p}_u \geq \bar{p}_v$ holds for node $u$ and $v$ such that $l_u \leq l_v$ and $u, v \neq q$.*

Lemma 2 implies that the estimated proximity of a visited node can not be lower than that of an unvisited and unselected node on the same/lower layer. Therefore, if the estimated proximity of a visited node is lower than the $K$-th highest proximity of the candidate nodes, all other unvisited and unselected nodes have lower proximities than the $K$-th highest proximities of the candidate nodes. Thus we can safely terminate the search process.

### 4.3.3 Incremental computation

As described in Section 4.3.2, by Definition 1, $O(n)$ time is required to compute the estimated proximity for each node. In this section, we show our approach to efficiently compute the estimated proximity. We assume that node $u$ is visited and selected immediately after node $u'$ in the search process. In other words, we visit and select these nodes in order $u'$ and $u$. In this section, let $\bar{p}_{u,1}$, $\bar{p}_{u,2}$, and $\bar{p}_{u,3}$ be first, second, and third terms in Equation (8), respectively. That is, $\bar{p}_u = c'(\bar{p}_{u,1} + \bar{p}_{u,2} + \bar{p}_{u,3})$.

We compute the estimation of $u$ as follows:

**Definition 2** (INCREMENATAL UPDATE). *For the given graph and query node, if $u' \neq q$, we compute the first, second, and third terms of the estimation of node $u$ from those of $u'$ in the search process as follows:*

$$\bar{p}_{u,1} = \begin{cases} \bar{p}_{u',1} & if\ l(u) = l(u') \\ \bar{p}_{u',2} + p_{u'} A_{max}(u') & otherwise \end{cases}$$
$$\bar{p}_{u,2} = \begin{cases} \bar{p}_{u',2} + p_{u'} A_{max}(u') & if\ l(u) = l(u') \\ 0 & otherwise \end{cases} \quad (9)$$
$$\bar{p}_{u,3} = (\bar{p}_{u',3}/A_{max} - p_{u'}) A_{max}$$

*If $u' = q$, $\bar{p}_{u,1} = p_q A_{max}(q)$, $\bar{p}_{u,2} = 0$, and $\bar{p}_{u,3} = (1 - p_q) A_{max}(u)$.*

We provide the following lemma for the incremental computation in the search process:

**Lemma 3** (INCREMENATAL UPDATE). *For node $u$, the estimated proximity can be exactly computed at the cost of $O(1)$ time by Definition 2.*

This property enables K-dash to efficiently compute the estimated proximity in the search process.

## 4.4 Search algorithm

Even though a detailed search algorithm of K-dash is described in Algorithm 4 in Appendix B.2, we outline our search algorithm as follows to make the paper self-contained:

1. We construct a breadth-first search tree rooted at the query node.
2. We visit a node in ascending order of tree layer and compute its estimated proximity by Definition 2.
3. If the estimated proximity of the visited node is not lower than the $K$-th proximity of the candidate nodes, the node can be an answer node. Therefore, we compute the proximity of the node and return to step 2.
4. Otherwise, we terminate the search process since the node and other unselected nodes can not be answer nodes (Lemma 2).

## 5. THEORETICAL ANALYSIS

This section provides a theoretical analysis that confirms the accuracy and complexity of K-dash. Let $n$ be the number of nodes. Proofs of each theorem in this section are shown in Appendix A.

We show that K-dash finds the top-k highest proximity nodes accurately (without fail) as follows:

**Theorem 2** (EXACT SEARCH). *K-dash guarantees the exact answer in finding the top-k highest proximity nodes.*

We discuss the complexity of the existing approximate algorithm B_LIN and NB_LIN [22] and then that of K-dash.

**Theorem 3** (THE APPROXIMATE ALGORITHM). *B_LIN and NB_LIN both require $O(n^2)$ space and $O(n^2)$ time to find the top-k highest proximity nodes.*

**Theorem 4** (SPACE COMPLEXITY OF K-DASH). *K-dash requires $O(n^2)$ space to find the top-k highest proximity nodes.*

**Theorem 5** (TIME COMPLEXITY OF K-DASH). *K-dash requires $O(n^2)$ time to find the top-k highest proximity nodes.*

Theorems 3, 4, and 5 show that K-dash has, in the worst case, the same space and time complexities as the previous approximate approaches. However, the space and time complexities of K-dash is, in practice, $O(n+m)$, which are smaller than those of the previous approximate approaches. This is because the number of non-zero elements in the inverse matrices is $O(m)$ as shown in the next section. In the next section, we confirm the effectiveness of our approaches by presenting the results of extensive experiments.

## 6. EXPERIMENTAL EVALUATION

We performed experiments to demonstrate K-dash's effectiveness in a comparison to NB_LIN by Tong et al. [22] and Basic Push Algorithm by Gupta et al. [7]. NB_LIN was selected since, as reported in [22], it outperforms the iterative approach and the approximation approach by Sun et al. [19], and it yields similar results to B_LIN, which was also proposed by by Tong et al., in all of our datasets. NB_LIN has one parameter: the target rank of the low-rank approximation. We used SVD as the low-rank approximation as proposed by Tong et al. Note that NB_LIN can compute proximities quickly at the expense of exactness. Basic Push Algorithm is an approach that can find top-k nodes efficiently for Personalized PageRank. The definitions of RWR and Personalized PageRank are very similar [6]. Even though Avrachenkov et al. also proposed an efficient approach for Personalized PageRank based top-k search [2], we compared K-dash to Basic Push Algorithm. This is because Basic Push Algorithm uses precomputed proximities of hub nodes to estimate the upper bounding proximities [7]; this implies Basic Push Algorithm theoretically guarantees that the recall of its answer result is always 1 while the approach of Avrachenkov et al. does not. Basic Push Algorithm is an approximate approach and the number of answer nodes yielded by this approach can be more than $K$.

Our experiments will demonstrate that:

- Efficiency: K-dash can outperform the approximate approach by several order of magnitude in terms of search time for the real datasets tested (Section 6.1).
- Exactness: Unlike the approximate approach, which sacrifices accuracy, K-dash can find the top-k nodes exactly (Section 6.2).
- Effectiveness: The components of K-dash, sparse matrices computation and tree estimation, are very effective in identifying top-k nodes (Section 6.3).

The results of the application of K-dash to a real dataset are reported in Appendix D.

We used the following five public datasets in the experiments: *Dictionary*, *Internet*, *Citation*, *Social*, and *Email*. The details of datasets are reported in Appendix C. In this section, *K-dash* represents the results of finding the top five nodes with the hybrid reordering approach. We set the restart parameter, $c$, at 0.95 as in the previous works [22, 8]. We evaluated the search performance through wall clock time. All experiments were conducted on a Linux quad 3.33 GHz Intel Xeon server with 32GB of main memory. We implemented our algorithms using GCC.

### 6.1 Efficiency of K-dash

We assessed the search time needed for K-dash, NB_LIN, and Basic Push Algorithm. Figure 2 shows the results. The results of K-dash are referred to as *K-dash(K)*, where $K$ is the number of answer nodes. We set the target rank of SVD to 100 and 1,000 (referred to as *NB_LIN(100)* and *NB_LIN(1,000)*). Note that the number of answer nodes, $K$, has no impact in NB_LIN since it computes approximate proximity scores for all nodes. *BPA(K)* indicates the results of Basic Push Algorithm where $K$ is the number of answer nodes and the number of hub nodes is set to 1,000.

This figure shows that our method is much faster than the previous approaches under all conditions examined. Specifically, K-dash is more than 68,000 times faster than NB_LIN and 690,000 times faster than Basic Push Algorithm. As described in Section 5, NB_LIN takes $O(n^2)$ time to compute proximities. Even though K-dash theoretically requires $O(n^2)$ time as shown in Lemma 5, it does not, in practice, take $O(n^2)$ time to find the answer nodes. This is because the number of non-zero elements in the inverse matrices is $O(m)$ in practice as shown in Section 6.3.1. That is, the time complexity of K-dash is, in practice, $O(n+m)$, because it takes $O(n+m)$ time for breadth-first tree construction and

---
[6] In Personalized PageRank, a random particle returns to the start node set, not the start node.

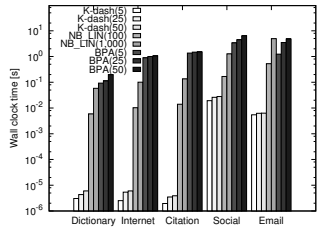 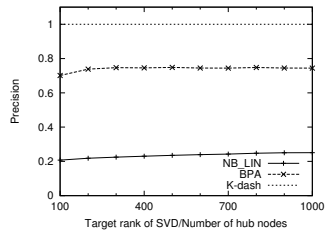 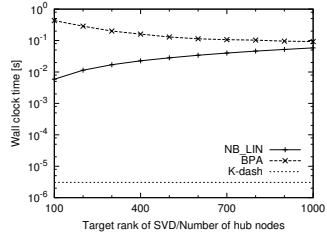 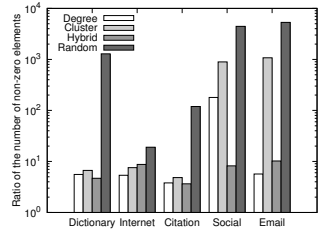

Figure 2: Efficiency of K-dash.

Figure 3: Precision of NB_LIN.

Figure 4: Wall clock time of NB_LIN.

Figure 5: Effect of reordering approaches.

$O(m)$ time for proximity computations. Therefore, our approach can find the answer nodes more efficiently than the previous approaches.

## 6.2 Exactness of the search results

One major advantage of K-dash is that it guarantees the exact answer, but this raises the following simple question: 'How successful are the approximate approaches in providing the exact answer even though it sacrifices exactness?'.

To answer this question, we conducted comparative experiments. We used precision as the metric of accuracy. Precision is the fraction of answer nodes among top-k results by each approach that match those of the original iterative algorithm. Figure 3 shows the precision and Figure 4 shows the wall clock time. In this experiment, we used various target rank setting and various number of hub nodes for NB_LIN and Basic Push Algorithm, respectively. We used Dictionary as the dataset in these experiments.

As we can see from Figure 3, the precision of K-dash is 1 because it finds the top-k nodes without fail. NB_LIN, on the other hand, has lower precision. Figure 4 shows that K-dash greatly reduces the computation time while it guarantees the exact answer. The efficiency of NB_LIN depends on the parameters used.

And Figure 3 and Figure 4 show that NB_LIN forces a trade-off between speed and accuracy. That is, as the target rank decreases, the wall clock time decreases but the precision decreases. NB_LIN does not guarantee that the answer results are accurate, and so can miss the exact top-k nodes. K-dash also computes estimate proximities, but unlike the approximate approach, K-dash does not discard any answer nodes in the search process.

Figure 3 shows that the precision of Basic Push Algorithm is almost constant against the number of hub nodes. Figure 4 indicates that the search speed of the approach increases as the number of hub nodes increases. This is because Basic Push Algorithm utilizes precomputed proximities of hub nodes to estimate the proximities of a query node. Figures 3 and 4 show that our approach is much faster than the previous approaches while guaranteeing exactness.

## 6.3 Effectiveness of each approach

In the following experiments, we examine the effectiveness of the two core techniques of K-dash: sparse matrices computation and tree estimation.

### 6.3.1 Reordering approach

K-dash utilizes the inverse matrices of lower/upper triangulars obtained by LU decomposition to compute the proximities of selected nodes in the search process. The number of non-zero elements in these inverse matrices is a factor that influences the search and memory cost. We take three approaches to reduce the number of non-zero elements as described in Section 4.2.2. Accordingly, we evaluated the ratio of the number of non-zero elements to that of edges in each reordering approach. Figure 5 shows the results. In this figure, *Random* represents the results achieved when nodes are arranged in random order.

As we can see from the figure, our approaches (Degree, Cluster, and Hybrid reordering) yield many fewer non-zero elements than random reordering. This figure also indicates that our approach makes the number of non-zero elements near to that of the edges of the given graph in all datasets if we adopt Hybrid reordering approach. That is, the space complexity of K-dash is $O(m)$. Owing to the small size of the inverse matrices, K-dash achieves excellent search performance as shown in Figures 2.

### 6.3.2 Precomputation time

Our approach uses the inverse matrices of lower/upper triangulars in the search process. That is, these matrices must be computed in the precomputing process. Figure 6 shows the process time in the precomputing process.

Figure 6 indicates that our reordering approach can enhance the process time; it is up to 140 times faster than the Random reordering approach. There are two reasons for this result. The first is that the inverse matrices have a sparse data structure as shown in Figure 5. The second is that elements of the inverse matrices which correspond to cross partition edges between 1st to $\kappa$-th partition are zero due to Equation (4), (5), (6), and (7) [7]. Therefore we can effectively skip the computations of these elements. As a result, we can efficiently compute the inverse matrices. Additional experiments confirmed that our approach needs less precomputation time due to its sophisticated sparse data structure than the other approaches. For example, NB_LIN needs several weeks to compute SVD because SVD requires $O(n^3)$ time, while our approach needs several hours.

### 6.3.3 Tree estimation

As mentioned in Section 4.3, K-dash skips unnecessary proximity computations in the search process. To show the effectiveness of this idea, we removed the pruning technique from K-dash, and reexamined the wall clock time. Figure 7 shows the result. K-dash without the pruning technique is abbreviated to *Without pruning* in this figure.

---

[7] For Dictionary, the improvement yielded by our approach was marginal. This is because the Louvain Method divides this dataset into one large partition and many small partitions which limits the effectiveness of our approach.

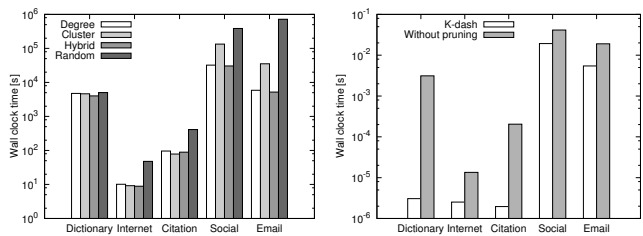

**Figure 6: Precomputation time.**  **Figure 7: Effect of tree estimation.**

The results show that the pruning technique can provide efficient search for all datasets used; these results indicate that our approach is effective for various edge distributions. K-dash is up to $1,020$ times faster if the pruning method is used. This is because we can effectively terminate the search process with the estimation technique. These results (compare Figure 2 to Figure 7) also show that, by , K-dash can find the top-k nodes faster than NB_LIN even if K-dash computes the proximities of all nodes. To evaluate the effectiveness of our approach for various proximity distributions, we subjected it to additional evaluations using various values of the restart probability $c$. The results confirmed that our approach can efficiently find the top-k nodes under all conditions examined; we can effectively prune unnecessary proximity computations.

## 7. CONCLUSIONS

This paper addressed the problem of finding the top-k nodes for a given node efficiently. As far as we know, this is the first study to address the top-k node search problem with the guarantee of exactness. Our proposal, K-dash, is based on two ideas: (1) It computes the proximities of selected nodes efficiently by use of inverse matrices, and (2) It skips unnecessary proximity computations in finding the top-k nodes, which greatly improves overall efficiency. Our experiments show that K-dash works as expected; it can find the top-k nodes at high speed; specifically, it is significantly faster than existing approximate methods. Top-k search based on RWR is fundamental for many applications in various domains such as image captioning, recommender systems, and link prediction. The proposed solution allows the top-k nodes to be detected exactly and efficiently, and so will help to improve the effectiveness of future applications.

# APPENDIX

## A. PROOFS

In this section, we show the proofs of all lemmas and theorems in this paper.

### A.1 Theorem 1

**Proof.** We prove the theorem by a reduction from the ELIMINATION ORDERING problem [23]. An instance of the ELIMINATION ORDERING problem consists of a graph, node elimination ordering, and the chordal graph that can be obtained by the graph and the elimination ordering. Given the graph, the ELIMINATION ORDERING problem finds the minimum number of edges whose addition makes the graph chordal by changing node ordering.

We transform an instance of the ELIMINATION ORDERING problem to an instance of the INVERSE MATRICES problem as follows: for the graph of the ELIMINATION ORDERING problem, we create matrix $\mathbf{A}$. That is, we make the adjacency-list matrix from the graph. For the node elimination ordering, we create node ordering, and we create matrix $\mathbf{L}^{-1}$ / $\mathbf{U}^{-1}$ for the chordal graph.

Given this mapping, it is easy to show that there exists a solution to the ELIMINATION ORDERING problem with the minimum number of edge additions if and only if there exists a solution to the INVERSE MATRICES problem with the minimum increase of non-zero elements in the inverse matrices. The INVERSE MATRICES problem is trivial in **NP**. □

### A.2 Lemma 1

**Proof.** If node $u$ is not the query node, the following equation holds from Equation (1):

$$p_u = (1-c)(A_{u1}p_1 + A_{u2}p_2 + \ldots + A_{uu}p_u + \ldots + A_{un}p_n)$$

Since more than two upper/lower layer nodes can not be directly connected to node $u$ in the breadth-first search tree, if the set of directly neighboring nodes (adjacent nodes) of node $u$ is $N_u$, $p_u$ is represented as follows:

$$p_u = c' \sum_{v \neq u} A_{uv} p_v = c' \sum_{v \in N_u} A_{uv} p_v$$

$$\leq c' \left\{ \sum_{v \in \{V_{l_u-1}(u) + V_{l_u}(u)\}} A_{uv} p_v + \sum_{v \in V \setminus V_s} A_{uv} p_v \right\}$$

Since $p_v$ is probability, $\sum_{v \in V \setminus V_s} p_v = 1 - \sum_{v \in V_s} p_v$ holds. Therefore,

$$p_u \leq c' \left\{ \sum_{v \in V_{l_u-1}(u)} p_v A_{max}(v) + \sum_{v \in V_{l_u}(u)} p_v A_{max}(v) + \left(1 - \sum_{v \in V_s} p_v\right) A_{max} \right\} = \bar{p}_u$$

If node $u$ is the query node, it is obvious $\bar{p}_u \geq p_u$ since $\bar{p}_u = 1$ and $0 \leq p_u \leq 1$. Thus the inequality holds. □

**Example.** Let node $u_1$ be a query node in a directed graph in Figure 8. As described in Section 4.3, node/layer numbers of all the nodes are assigned by breadth-first search tree; node $u_1$ forms layer 0, node $u_2$ and $u_3$ form layer 1, node $u_4$ and $u_5$ form layer 2, and node $u_6$ and $u_7$ form layer 3. And we assume that we visit and select nodes in ascending order of their node number. For node $u_5$, the following equation holds from Equation (1) since $A_{51}, A_{53}, A_{57} = 0$:

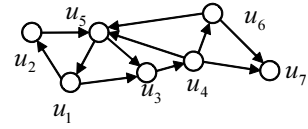

**Figure 8: An example graph.**

$$p_{u_5} = c'(A_{51}p_1 + A_{52}p_2 + A_{53}p_3 + A_{54}p_4 + A_{56}p_6 + A_{57}p_7)$$
$$= c'(A_{52}p_2 + A_{54}p_4 + A_{56}p_6)$$

Since the proximities of node $u_1$, $u_2$, $u_3$ and $u_4$ are already computed before node $u_5$ and node $u_6$, the following equation holds:

$$p_{u_5} \leq c'(p_2 A_{max}(u_2) + p_4 A_{max}(u_4) + (1-p_1-p_2-p_3-p_4)A_{max}) = \bar{p}_{u_5}$$

Note that our estimation approach takes into account edges of selected nodes and unvisited nodes as $A_{max}(u)$ and $A_{max}$, respectively. For example, non-tree edges $A_{54}$ and $A_{56}$ are taken as $A_{max}(u_4)$ and $A_{max}$, respectively.

### A.3 Lemma 2

**Proof.** If $l_u = l_v$, it is obvious that $\bar{p}_u = \bar{p}_v$. If $l_u = l_v - 1$, the following inequality holds since $V_{l_v}(v) = \emptyset$:

$$\bar{p}_v = c' \left\{ \sum_{w \in V_{l_u}(u)} p_w A_{max}(w) + \left(1 - \sum_{w \in V_s} p_w\right) A_{max} \right\} \leq \bar{p}_u$$

And if $l_u \geq l_v - 2$, the following inequality similarly holds since $V_{l_v}(v) = \emptyset$ and $V_{l_v-1}(v) = \emptyset$:

$$\bar{p}_v = c' \left(1 - \sum_{w \in V_s} p_w\right) A_{max} \leq \bar{p}_u$$

which completes the proof. □

### A.4 Lemma 3

**Proof.** We first prove that, if $u' = q$, $\bar{p}_u$ can be exactly computed from $\bar{p}_{u,1}$, $\bar{p}_{u,2}$, and $\bar{p}_{u,3}$. In this case, $V_{l_u-1}(u) = q$, $V_{l_u}(u) = 0$, and $V_s = q$. Therefore, it is obvious that $c'(\bar{p}_{u,1} + \bar{p}_{u,2} + \bar{p}_{u,3}) = \bar{p}_u$ holds by Definition 2.

We next prove that, if $u' \neq q$, the estimate proximity of node $u$ can be exactly computed from that of node $u'$.

If $l_u = l_{u'}$, $V_{l_u-1}(u) = V_{l_{u'}-1}(u')$ and $V_{l_u}(u) = V_{l'_u}(u') + u'$. Therefore,

$$\bar{p}_{u,1} - \bar{p}_{u',1} = \sum_{v \in \{V_{l_u-1}(u) - V_{l_{u'}-1}(u')\}} p_v A_{max}(v) = 0$$

and

$$\bar{p}_{u,2} - \bar{p}_{u',2} = \sum_{v \in \{V_{l_u}(u) - V_{l'_u}(u')\}} p_v A_{max}(v) = p_{u'} A_{max}(u')$$

Otherwise (i.e. $l_u = l_{u'} + 1$), $V_{l_u-1}(u) = V_{l_{u'}}(u') + u'$ and $V_{l_u}(u) = \emptyset$. Therefore,

$$\bar{p}_{u,1} = \sum_{v \in \{V_{l_{u'}}(u') + u'\}} p_v A_{max}(v) = \bar{p}_{u',2} + p_{u'} A_{max}(u')$$

and

$$\bar{p}_{u,2} = \sum_{v \in \emptyset} p_v A_{max}(v) = 0$$

Since node $u$ is visited immediately after node $u'$,

$$(\bar{p}_{u',3} - \bar{p}_{u,3})/A_{max} = p_{u'}$$

Therefore, $\bar{p}_{u,1}$, $\bar{p}_{u,2}$, and $\bar{p}_{u,3}$ can be exactly computed from $\bar{p}_{u',1}$, $\bar{p}_{u',2}$, and $\bar{p}_{u',3}$, respectively.

We finally prove that it takes $O(1)$ time to compute $\bar{p}_{u,1}$, $\bar{p}_{u,2}$, and $\bar{p}_{u,3}$ in the search process. If $u' = q$, $\bar{p}_{u,1}$, $\bar{p}_{u,2}$, and $\bar{p}_{u,3}$ are defined by Definition 2. As described in Section 4.3.1, both $A_{max}$ and $A_{max}(u)$ can be precomputed. $\bar{p}_{u',1}$, $\bar{p}_{u',2}$, $\bar{p}_{u',3}$, and $p_{u'}$ are already computed before computing $\bar{p}_u$ in the search process if $u' \neq q$. This completes the proof. □

## A.5 Theorem 2

**Proof.** Let $\theta$ be the $K$-th highest proximity among the candidate nodes in the search process. And let $\theta_K$ be the $K$-th highest proximity among the answer nodes (i.e. $\theta_K$ is the lowest proximity among the answer nodes).

We first prove that $\theta$ is monotonic non-decreasing in the search process of K-dash. To find the answer nodes in the search process, we first set $\theta$ at 0 and set the dummy nodes as the candidates. We maintain the candidate as the best result; when we find a node with higher proximity, its proximity is greater than $\theta$, the candidate is replaced by the new node (see Algorithm 4). This makes $\theta$ higher. Therefore, $\theta$ keeps increasing in the search process.

In the search process, since $\theta$ is monotonic non-decreasing and $\theta_K \geq \theta$, the estimate proximities of answer nodes are never lower than $\theta$ (Lemma 1). The algorithm discards a node if (1) its estimated proximity is lower than $\theta$, or (2) its upper/same layer unselected node has estimated proximity lower than $\theta$. Since the estimated proximity of a node can not be lower than that of a node on the same or lower layer (Lemma 2), the answer nodes can never be pruned during the search process. □

## A.6 Theorem 3

**Proof.** We first prove that B_LIN and NB_LIN [22] both need $O(n^2)$ space. The off-line process of B_LIN first partitions the adjacency matrix by METIS [9], and then decomposes the matrix into the within-partition edge matrix and the cross-partition edge matrix. It next performs low-rank approximation for the cross-partition edge matrix and obtains two orthogonal matrices and one diagonal matrix. It then computes the product of the within-partition edge matrix and the orthogonal matrices.

The off-line process of NB_LIN first performs low-rank approximation for the adjacency matrix and obtains two orthogonal matrices and one diagonal matrix. It then computes the product of these matrices.

Both B_LIN and NB_LIN hold the matrix product and two orthogonal matrices to compute the proximities. The matrix product and orthogonal matrices have size of $O(n^2)$. Therefore, B_LIN and NB_LIN both require $O(n^2)$ space.

Next, we prove that B_LIN and NB_LIN both need $O(n^2)$ time. They compute the proximities of nodes by multiplying the vector $\mathbf{q}$, the matrix product, and orthogonal matrices. Even though the size of the vector $\mathbf{q}$ is $O(n)$, that of the matrix product and orthogonal matrices is $O(n^2)$. Therefore, B_LIN and NB_LIN both require $O(n^2)$ time. □

---

**Algorithm 1** Degree reordering
**Input:** $\mathbf{A}$, the column normalized adjacent matrix
**Output:** $\mathbf{A}'$, the reordered matrix of $\mathbf{A}$
1: arrange nodes in ascending order of their degrees;
2: compute matrix $\mathbf{A}'$ by interchanging the rows and columns of matrix $\mathbf{A}$ by the degree order;
3: **return** $\mathbf{A}'$;

---

**Algorithm 2** Cluster reordering
**Input:** $\mathbf{A}$, the column normalized adjacent matrix
**Output:** $\mathbf{A}'$, the reordered matrix of $\mathbf{A}$
1: divide nodes into $\kappa$ partitions $P_1, P_2, \ldots, P_\kappa$ by Louvain method;
2: create new partition $P_{\kappa+1} = \emptyset$;
3: **for** $i := 1$ to $\kappa$ **do**
4:   remove nodes whose edges cross more than two partitions from partition $P_i$;
5:   append the removed nodes to $P_{\kappa+1}$;
6: **end for**
7: compute matrix $\mathbf{A}'$ by interchanging the rows and columns of matrix $\mathbf{A}$ by the partitions;
8: **return** $\mathbf{A}'$;

---

## A.7 Theorem 4

**Proof.** To compute the estimation, K-dash holds the maximum elements of the matrix $\mathbf{A}$, the previous estimated proximity, and the previous proximity. It needs $O(n)$ space to hold these values. K-dash keeps the inverse matrices to compute the proximities. The number of non-zero elements of these matrices is $O(n^2)$ in the worst case. Therefore, it requires $O(n^2)$ space to keep the inverse matrices. Therefore, our approach requires $O(n^2)$ space to find top-k highest proximity nodes. □

## A.8 Theorem 5

**Proof.** To find the answer nodes, K-dash first constructs the breadth-first search tree, and computes the estimated proximity of the visited node. It next computes the proximity of the node if the node is not pruned by the estimation. K-dash needs $O(n+m)$ time to construct the tree and $O(n)$ time if it can not prune any nodes by the estimation. This is because it takes $O(1)$ time to compute the estimation for each node (Lemma 3). K-dash needs $O(n^2)$ times to compute the proximities of all nodes since the inverse matrices have size of $O(n^2)$ in the worst case. So K-dash needs $O(n^2)$ time to find the top-k highest proximity nodes. □

## B. ALGORITHMS

In this section, we show the algorithms for reordering approaches and K-dash.

## B.1 Reordering approach

We interchange the rows and columns of matrix $\mathbf{A}$ to reduce the number of non-zero elements in the inverse matrices. Since the INVERSE MATRICES problem is **NP**-complete, we take three approximation solutions to the problem: degree reordering, cluster reordering, and hybrid reordering.

Algorithm 1 depicts our degree reordering approach. This approach reduces non-zero elements by arranging low degree nodes to the upper/left elements in matrix $\mathbf{A}$. It first computes the degrees of all nodes and arranges the nodes according to their degrees (line 1). It then computes the reordered matrix by the degree order (line 2).

**Algorithm 3** Hybrid reordering

**Input:** $\mathbf{A}$, the column normalized adjacent matrix
**Output:** $\mathbf{A}'$, the reordered matrix of $\mathbf{A}$
1: divide nodes into $\kappa + 1$ partitions $P_1, P_2, \ldots, P_{\kappa+1}$ by cluster reordering;
2: **for** $i := 1$ to $\kappa + 1$ **do**
3:   arrange nodes in the $i$-th partition in ascending order of their degrees;
4: **end for**
5: compute matrix $\mathbf{A}'$ by interchanging the rows and columns of matrix $\mathbf{A}$ by the partitions and the degree order;
6: **return** $\mathbf{A}'$;

**Algorithm 4** K-dash

**Input:** $q$, the query node
  $K$, the number of answer nodes
  $\mathbf{L}^{-1}$, the inverse matrix of $\mathbf{L}$
  $\mathbf{U}^{-1}$, the inverse matrix of $\mathbf{U}$
**Output:** $V_a$, the set of answer nodes
1: $\theta = 0$;
2: $V_s = \emptyset$;
3: $V_a = \emptyset$;
4: append $K$ dummy nodes to $V_a$;
5: compute the breadth-first search tree of node $q$;
6: **while** $V_s \neq V$ **do**
7:   $u := \text{argmin}(l_v | v \in V \setminus V_s)$;
8:   compute the estimate proximity of node $u$, $\bar{p}_u$;
9:   **if** $\bar{p}_u < \theta$ **then**
10:    **return** $V_a$;
11:  **else**
12:    compute the proximity, $p_u$, by $\mathbf{L}^{-1}$ and $\mathbf{U}^{-1}$;
13:    **if** $p_u > \theta$ **then**
14:      $v := \text{argmin}(p_w | w \in V_a)$;
15:      remove node $v$ from $V_a$;
16:      append node $u$ to $V_a$;
17:      $\theta := \min(p_w | w \in V_a)$;
18:    **end if**
19:  **end if**
20:  append node $u$ to $V_s$;
21: **end while**
22: **return** $V_a$;

We show our cluster reordering approach in Algorithm 2. It reduces non-zero elements in the inverse matrices by transferring nodes whose edges cross partitions into the $\kappa+1$-th partition. It first partitions the graph into $\kappa$ partitions by Louvain method (line 1). It checks each node as to whether the node has any cross-partition edges. If the node has cross-partition edges, it transfers the node to $\kappa + 1$-th partition (lines 3-6). It finally computes the reordered matrix by the partitions (line 7).

Algorithm 3 shows our hybrid reordering approach. It combines the above two approaches. It first obtains the reordered matrix by the cluster reordering approach (line 1). It then arranges nodes in each partition by their degrees (line 2-4). It finally computes the reordered matrix by the partitions and the degree order (line 5).

### B.2 Search algorithm

Our main approach to finding the answer nodes is to compute the proximities of selected nodes by the inverse matrices, and to use the estimated proximities to skip unnecessary proximity computations.

Algorithm 4 shows the search algorithm that efficiently finds $K$ highest proximity nodes for the query node. In this algorithm, $\theta$ and $V_a$ indicate the $K$-th highest proximity among the candidate nodes and the set of candidate/answer nodes, respectively.

In the search process, K-dash first sets the candidate nodes by appending $K$ dummy nodes where the proximities of the dummy nodes are all 0 (line 4), it then constructs the breadth-first search tree (line 5). K-dash then visits nodes according to the tree layer one by one (line 7), and computes the estimated proximity of each node (line 8). If the estimated proximity of the visited node is lower than $\theta$, the node cannot be the answer node (Lemma 1), and the proximities of all other unselected nodes cannot be higher than $\theta$ (Lemma 2). Therefore it terminates the search process (lines 9-10). Otherwise, the visited node may be an answer node. Thus it computes the proximity of the node (line 12). If the computed proximity is higher than $\theta$, it updates the candidate set, $V_a$, and $\theta$ (lines 13-18). It returns the candidate set, $V_a$, as the answer nodes set (line 22).

As shown in Algorithm 4, this algorithm automatically terminates the process if the estimated proximity is lower than $\theta$. That is, this algorithm does not require any user-defined inner-parameters.

### C. EXPERIMENTAL DATASETS

We used the following five public datasets:

- Dictionary [8]: This dataset was taken from word network in FOLDOC [9]. FOLDOC is a famous on-line dictionary of computing subjects. An edge from node $u$ to node $v$ exists in the graph if and only if in the FOLDOC dictionary term $v$ is used to describe the meaning of term $u$. The number of nodes and edges are $13,356$ and $120,238$, respectively.

- Internet [10]: We used a snapshot of the structure of the Internet at the level of autonomous systems. This graph was constructed from BGP tables posted by the University of Oregon Route Views Project [11]. Oregon Route Views Project allows Internet users to view global BGP routing information from the perspective of other locations around the Internet. The number of nodes and edges are $22,963$ and $48,436$, respectively.

- Citation [12]: This graph is weighted network of co-authorships between scientists posting preprints on the Condensed Matter E-Print Archive [13]. The Condensed Matter E-Print Archive is the fully automated e-print archive for condensed matter preprints which is a specialized field in physics. The number of nodes and edges are $31,163$ and $120,029$, respectively.

- Social [14]: This graph is taken from Epinions.com [15]. Epinions.com is a general consumer review site. Members of the site can decide whether to trust each other. This dataset is who-trust-whom online social network which has $131,828$ nodes and $841,372$ edges.

- Email [16]: The graph was generated using email data from a large European research institution. In this graph, each node corresponds to an email address. And a directed edge between nodes $u$ and $v$ represents user of address $u$ sent at least one message to address $v$. This dataset has $265,214$ nodes and $420,045$ edges.

---

[8] http://vlado.fmf.uni-lj.si/pub/networks/data/dic/foldoc/foldoc.zip
[9] http://foldoc.org/
[10] http://www-personal.umich.edu/ mejn/netdata/as-22july06.zip
[11] http://routeviews.org/
[12] http://www-personal.umich.edu/ mejn/netdata/cond-mat-2003.zip
[13] http://arxiv.org/archive/cond-mat
[14] http://snap.stanford.edu/data/soc-sign-epinions.html
[15] http://www.epinions.com/
[16] http://snap.stanford.edu/data/email-EuAll.html

Table 2: Ranked lists by K-dash and NB_LIN for company and operating system names.

| Term | Method | Rank | | | | |
|---|---|---|---|---|---|---|
| | | 1 | 2 | 3 | 4 | 5 |
| Microsoft | K-dash | Microsoft | MS-DOS | IBM PC | Microsoft Windows | Microsoft Corporation |
| | NB_LIN | Microsoft | microsecond | CS-Prolog | MICRO SAINT | Microsoft Basic |
| APPLE | K-dash | APPLE | Apple Attachment Unit Interface | Apple II | Apple Computer, Inc. | APPC |
| | NB_LIN | APPLE | APIC | Personalized Array Translator | I-APL | CEEMAC+ |
| Microsoft Windows | K-dash | Microsoft Windows | W2K | Windows/386 | Windows 3.0 | Windows 3.11 |
| | NB_LIN | Microsoft Windows | Microsoft Networking | Microsoft Network | W2K | Thumb |
| Mac OS | K-dash | Mac OS | Macintosh user interface | Macintosh file system | multitasking | Macintosh Operating System |
| | NB_LIN | Mac OS | Rhapsody | SORCERER | Macintosh Operating System | PowerOpen Association |
| Linux | K-dash | Linux | Linux Documentation Project | Unix | lint | Linux Network Administrators' Guide |
| | NB_LIN | Linux | Linux Documentation Project | SL5 | debianize | SLANG |

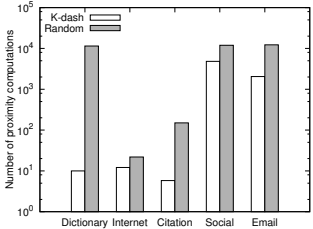

Figure 9: Comparison of root node selection.

## D. ADDITIONAL EXPERIMENTS

In this section, we show the result of additional experiment on root node selection and case-studies on two company names and three operating system names.

### D.1 Root node selection

Our estimate algorithm sets the query node as the root node to find the top-k nodes efficiently. To show the effectiveness of this idea, we show the number of proximity computations in Figure 9. In this figure, *Random* represents the case of selecting the root node at random.

Our root node selection method requires fewer proximity computations than the random methods. In the search process, we first compute the breadth-first search tree from the root nodes. As the query node and its neighboring nodes have high proximity, we can obtain high proximity nodes with this approach. As a result, we can more effectively estimate the proximities of unselected nodes.

### D.2 Case-studies

In this section, we show results of experimental case-studies on two company names and three operating system names to show the effectiveness of K-dash. In this experiment, we identified the high proximity terms for 'Microsoft', 'APPLE', 'Microsoft Windows', 'Mac OS', and 'Linux'. We set the target rank of SVD to 1,000 for NB_LIN. Table 2 shows the results. We omitted the results of Basic Push Algorithm due to space limitations.

The results of our method for the two company names make sense while those of the approximate method do not. For example, K-dash successfully detected the formal names of the two companies, 'Microsoft Corporation' for 'Microsoft' and 'Apple Computer, Inc.' for 'APPLE'. K-dash detected 'IBM PC' as the third-relevant term for 'Microsoft'. This result may seem strange, however, it is very reasonable if we consider the close relation of the companies. Microsoft was founded in 1975 by Bill Gates. In 1980, IBM chose Microsoft to supply the operating system for the IBM PC. As a result, Microsoft eventually became the leading vendor. For Apple, K-dash finds APPLE II as the third-relevant term which is an 8-bit PC of the company. APPLE II was invented by Steve Wozniak, who is co-founder of Apple with Steve Jobs, and was very popular from about 1980 until the first several years of MS-DOS. However, the approximate approach has difficulty in obtaining these intuitive results.

The results of K-dash for the three operating system names reveal that these operating systems have distinctive characteristics. The results of 'Microsoft Windows' are 'Microsoft Windows', 'W2K', 'Windows/386', 'Windows 3.0', and 'Windows 3.11'. All are Microsoft operating systems. These results reflect Microsoft's dominant market position. On the other hand, the results of 'Mac OS' include several of Apple's unique technical terms. For example, K-dash detects 'Macintosh user interface' as the second-relevant term for 'Mac OS'. Macintosh user interface is the graphical user interface used by Apple Computer's Macintosh family of PCs. The original Macintosh was the first successful PC to use a graphical user interface devoid of a command line. 'Macintosh file system', the third-relevant term for 'Mac OS', is Apple's disk file system adopted only in Mac OS. These results reflect Apple's culture of creativity. The results of 'Linux' imply its open culture. Unlike Microsoft Windows, the development of Linux is an example of free and open source software collaboration. Therefore, there are many projects that support the development of Linux. Linux Documentation Project is an all-volunteer project that maintains a large collection of GNU and Linux-related documentation and publishes the collection on-line.

In conclusion, the results of K-dash are strong indicative of the characteristics of each company and operating system. The approximate approach is ineffective in finding such high relevant terms for the two company names and the three operating system names.